\documentclass[aps,prl,twocolumn,superscriptaddress,showpacs]{revtex4}
\usepackage{graphicx}
\usepackage{dcolumn}
\usepackage{color}
\usepackage{mathrsfs}
\usepackage[breaklinks,colorlinks = true,linkcolor = blue,urlcolor=blue,citecolor=blue]{hyperref}

\begin{document}
\title{Observation of the novel type of ordering: Spontaneous ferriquadrupolar order}

\author{A.A. Zvyagin}
\affiliation{Max-Planck-Institut f\"ur Physik komplexer Systeme, Noethnitzer
Str., 38, D-01187, Dresden, Germany}
\affiliation{B.I.~Verkin Institute for Low Temperature Physics and 
Engineering of the National Academy of Sciences of Ukraine,
Nauky  Ave., 47, Kharkiv, 61103, Ukraine}
\affiliation{V.N.~Karazin Kharkiv National University, Svobody sq., 4 Kharkiv 61002 Ukraine} 
\author{K. Kutko} 
\affiliation{B.I.~Verkin Institute for Low Temperature Physics and 
Engineering of the National Academy of Sciences of Ukraine,
Nauky  Ave., 47, Kharkiv, 61103, Ukraine}
\author{D. Kamenskyi}
\affiliation{High Field Magnet Laboratory (HFML – EMFL), Radboud University, Toernooiveld 7, 6225 ED Nijmegen, The Netherlands}
\affiliation{FELIX Laboratory, Institute for Molecules and Materials, Radboud University, Toernooiveld 7c, 6225 AJ Nijmegen, The Netherlands}
\author{A.V. Peschanskii}
\affiliation{B.I.~Verkin Institute for Low Temperature Physics and 
Engineering of the National Academy of Sciences of Ukraine,
Nauky  Ave., 47, Kharkiv, 61103, Ukraine}
\author{S. Poperezhai}
\affiliation{B.I.~Verkin Institute for Low Temperature Physics and 
Engineering of the National Academy of Sciences of Ukraine,
Nauky  Ave., 47, Kharkiv, 61103, Ukraine}
\author{N.M. Nesterenko}
\affiliation{B.I.~Verkin Institute for Low Temperature Physics and 
Engineering of the National Academy of Sciences of Ukraine,
Nauky  Ave., 47, Kharkiv, 61103, Ukraine}

\begin{abstract}
Using Raman and infrared spectroscopies the spontaneous {\em ferriquadrupolar} ordering has been observed in the rare-earth-based system KDy(MoO$_4$)$_2$. Ordered quadrupoles in the electron subsystem attend non-equivalent distortions of rare-earth ions in the ordered phase. The mean field theory explaining the onset of such a type of ordering has been constructed.  

\end{abstract}

\date{\today}
\maketitle

Electronic correlations in condensed matter can change the behavior of electrons from the conventional metallic one to new, so called liquid states, such as electronic liquid crystal ones in metals, valence bond solids, or spin liquid states in insulating quantum spin systems \cite{liquid}. For example, in the latter the standard magnetic order (i.e., the ordering of magnetic dipoles) is suppressed down to the lowest temperatures due to the frustration of spin-spin interactions and/or enhanced quantum fluctuations in low-dimensional systems \cite{fr}. Individual magnetic moments (spins) remain disordered, while higher rank multipoles, first of all quadrupoles, that are caused by strong correlations between ions, can order under some special conditions. Unlike the order of magnetic dipoles, the quadrupolar order does not break the time-reversal symmetry. Quadrupolar ordering in insulating spin systems is often referred to as the ``spin nematic'' one, because that spin ordering is analogous to the known ordering of molecules in nematic phases of liquid crystals \cite{spnem}. For orbital electron moments the dipole moment of the orbital electron moment is frozen in most of compounds (including the compound KDy(MoO$_4$)$_2$ discussed below), and the orbital order reveals itself first of all in the charge ordering or in the ordering of quadrupolar moments.  The ``spin nematic'' order and the orbital/charge order breaks the rotational symmetry of electronic states. Strong nematic fluctuations were observed in Fe- and Cu-based superconductors \cite{Cusup} (nematicity is believed to be an essential property for Fe-based superconductors \cite{Fesup}), as well as in strontium ruthenates \cite{rut}, and high fractional Landau levels \cite{fracLL}. The quadrupolar ``hidden'' order is often difficult to detect experimentally because most of the available techniques is sensitive to dipole moments only. Also, dipole moments are coupled to the electromagnetic field, while there is no simple field, directly coupled to quadrupoles. 

The co-operative Jahn-Teller (JT) effect \cite{JT} gives the possibility to observe the quadrupolar ordering \cite{Ell,not}. Here quadrupolar orbital moments of the electron subsystem of a crystal interact with strains of the elastic subsystem, giving rise to the phase transition to the low-temperature phase, which is characterized by both, the quadrupolar electron ordering, and nonzero distortions of the oxigen polyhedrons around the JT ions. Rare-earth based JT compounds with the strong coupling between electron and elastic subsystems permit to observe the quadrupolar ``hidden'' order of several kinds. Mostly the ferroquadrupolar (by analogy with ferromagnetic) ordering (with a single sublattice of ordered orbital quadrupoles) was observed in JT systems \cite{JT}, however some rare-earth based compounds manifest the antiferroquadrupole ordering, where at least two sublattices of ordered quadrupoles exist at low temperatures, as in antiferromagnets \cite{Bor}. In this communication we report the first direct observation of the spontaneous {\em ferriquadrupolar} (with two non-equivalent sublattices of quadrupoles) ordering. Using the Raman and far infrared spectroscopy we observe clear features of the onset of two energetically non-equivalent ordered orbital quadrupolar moments in the rare-earth-based compound. That ordering is accompanied with the non-equivalent distortions of two groups of rare-earth ions, which reveal themselves in the onset of new electron optical lowest branches and new phonons, also observed in our experiments.    

KDy(MoO$_4$)$_2$ is the compound which represents the family of double molybdates with a general chemical formula MRe(MoO$_4$)$_2$, where M$^+$ is an alkali metal ion and Re$^{3+}$ is a rare-earth ion. The strong coupling between electronic excitations of Re$^{3+}$ ions and phonons together with the strong anisotropy of Re$^{3+}$ electronic states causes structural phase transitions, both spontaneous and induced by an externally applied magnetic field. Thus, the series of those compounds include materials which are perfect playground for studying the microscopic details of phase transitions related to the co-operative JT ones. In KDy(MoO$_4$)$_2$, previous investigations revealed the spontaneous phase transition at the temperature of order of 10~K \cite{Zvya}. The external magnetic field being applied in a certain direction suppresses the phase transition \cite{Leask}. In the high temperature phase KDy(MoO$_4$)$_2$ has the $Pbcn$($D^{14}_{2h}$) structure with $Z = 4$. At present low-temperature X-ray data are absent for KDy(MoO$_4$)$_2$, however the phenomenological approach shows that the phase transition leads to the doubling of the unit cell with the lowering of the symmetry to monoclinic \cite{Nest}. Nevertheless, the microscopic mechanism remains unclear. The series of studies \cite{Kharch} demonstrates that in the process of phase transformations two phase transitions take place with close critical temperatures. There is a data spread in the definition of the value of the critical temperatures \cite{Zvya,Leask,Kharch,Cooke,Mih,Vit,Kut,Zagv,Kolod,Pes}. 

Besides the question remains, how the distortions are ordered around non-equivalent ions of Dy$^{3+}$; there are four those ions in the high-temperature phase, and eight in the low-temperature ones. The study \cite{Pes} stresses, that the standard soft mode has not been detected. The lowest multiplet of the Dy$^{3+}$ ion, namely $^6H_{15/2}$, is split by the crystal field into eight doublets; the first excited level lies at 18~cm$^{-1}$ above the ground state; the next one is at about 80~cm$^{-1}$ \cite{suppl}. 

To investigate the temperature behavior of the lowest energy levels at temperatures 1.4$\div$25~K two spectroscopic methods, the far infrared (FIR) and Raman spectroscopy, have been used.The goal of our study is the determination of the type of ordering in the multi-sublattice JT system. 

Studied KDy(MoO$_4$)$_2$ samples were grown from a melt by slow cooling. The samples have the typical plate shape with the dimensions of 5$\times$5 mm$^2$ with the $b$-axis perpendicular to the plates. It allows for FIR transmission measurements with the polarizations of the electric component of the radiation $E_{\omega} \parallel a$ ($B_{3u}$) and $E_{\omega} \parallel c$ ($B_{1u}$). The FIR experiments have been performed using the Fourier-transform infrared spectrometer Bruker IFS113v. A mercury lamp has been used as a radiation source, and the liquid-helium-cooled silicon bolometer has been used as a detector. Raman scattering has been measured using the Nd:YAG (neodymium-doped yttrium aluminum garnet) solid-state laser ($\lambda_{exc}=532$~nm) with 36~mW incident power. The sample heating has been estimated by relative intensities of Stokes and anti-Stokes components of a phonon mode at 74~cm$^{-1}$. The scattered light has been collected in the 90 degree geometry, dispersed through a double stage monochromator Ramanor U-1000, and then detected with a cooled photomultiplier. 

The FIR spectra of KDy(MoO$_4$)$_2$ have a complicated structure. First of all, multiple reflections within the plane-parallel sample cause the Fabry-Perot type of modulation of the transmittance, which appears as fringes in the spectra. The periodicity of the fringes (see Fig. 1a) is determined by the thickness and the refractive index of the sample. The exact energies of the absorption modes were determined using the REFFIT script \cite{Kuz}, which also simulated the fringes. Second, at the same range of energies besides electronic excitations the spectra show the low energy phonons.

\begin{figure}
\begin{center}
\includegraphics[width=0.4\textwidth]{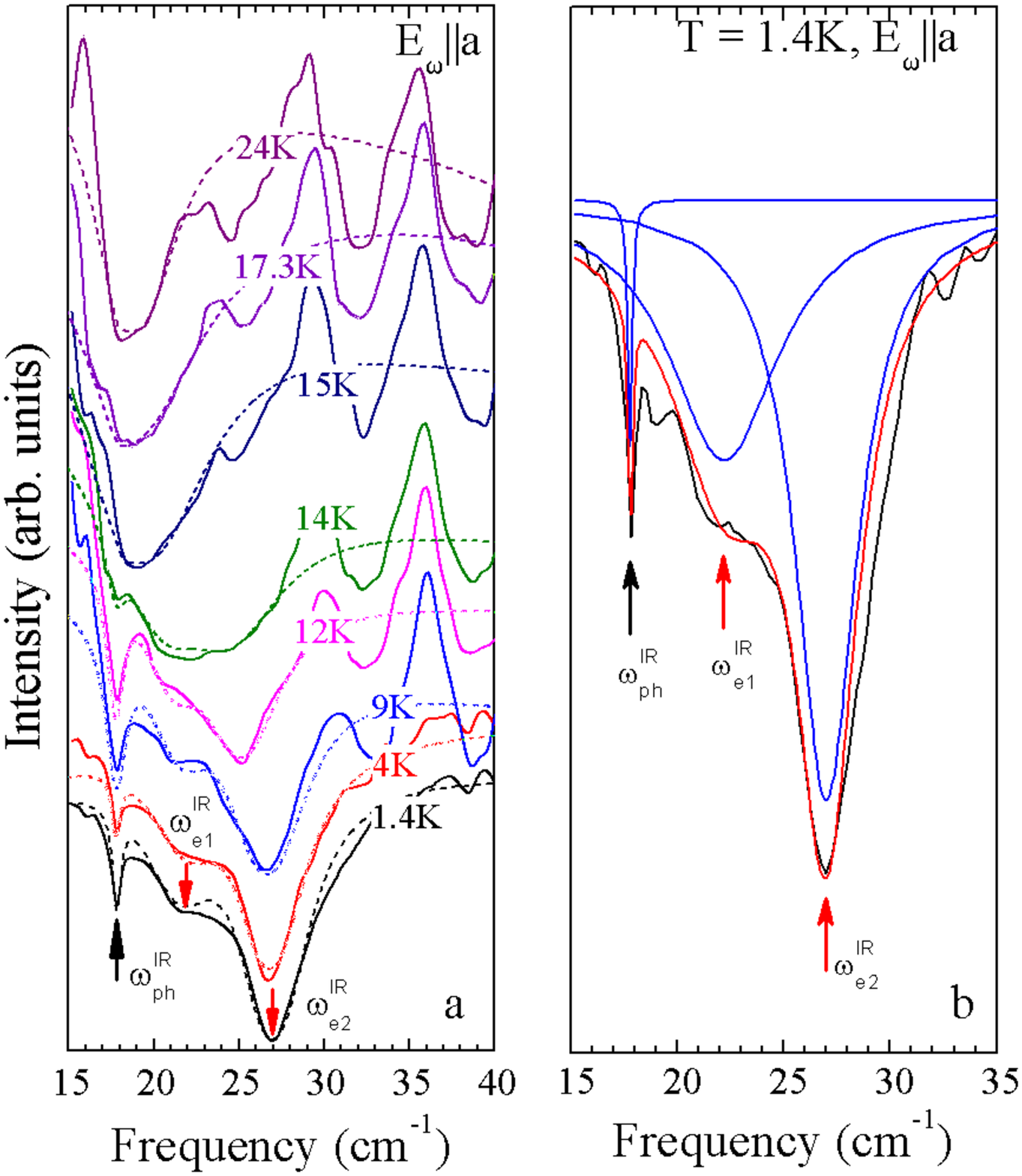}
\end{center}
\caption{(Color online) Left panel: The temperature evolution of the transmission (FIR) spectra of KDy(MoO$_4$)$_2$ for the $E_{\omega} \parallel a$ polarization. Black arrow shows the phonon mode $\omega_{ph}^{IR}$ and red arrows show the electron modes $\omega_{e1}^{IR}$ and  $\omega_{e2}^{IR}$. Solid lines: experiment; dashed lines: multiple reflections removed. Right panel: FIR transmission spectra for T=1.4~K; the black line describes the data of the experiment; three blue lines describe theoretical fitting within the Lorentz approximation for each of three observed lines; the red line is the sum of all fitting contributions.}
\label{fig1}
\end{figure}

Fig.~1 (left panel) displays the temperature evolution of the FIR spectra between 1.4 and 24~K for $E_{\omega} \parallel a$ polarization (experiments have been performed for other polarizations also). The lines in this temperature region have different half-widths and different temperature dependence of their energy positions. Fig 1 (right panel) shows the positions and spectral half-widths at 1.4~K; it was supposed that every line has the Lorentz profile. The most intensive line from the high energy side is determined as the first excited level of Dy$^{3+}$ ions; at higher temperatures (25$\div$18~K) its energy is about 18~cm$^{-1}$. By lowering the temperature to 1.5~K its position is changed to $\sim $27~cm$^{-1}$. Such a behavior was related \cite{Zvya,Leask} to the cooperative JT effect in KDy(MoO$_4$)$_2$, namely, to the antiferroquadrupolar ordering with two equivalent quadrupolar sublattices \cite{Leask}. The lowest energy excitation in this group is presented by the very narrow shape (its high-width is about 0.5~cm$^{-1}$). It is related \cite{Pop} to the vibrations of the layers  $\{$Dy$^{3+}$(MoO$_4$)$^{2-}\}^-$ as a whole. The energy and the half-width do not depend on the temperature at least in the temperature region where they are not overlapped with the wide electron line. Note, that at the external magnetic field  the electron line shifts to the higher energies \cite{suppl}. Meanwhile the phonon excitation close to 17.5~cm$^{-1}$ does not demonstrate a noticeable broadening compared with lower temperatures. At least, the weak electron excitation appears in the spectra near 22~cm$^{-1}$ at temperatures below about 15~K. We suppose that the onset of the new excitation is the result of the enlarging of the rhombic phase unit cell at least twice \cite{Pes}. The number of JT centers which are connected with Dy$^{3+}$ ions is twice as much as in the high-temperature phase. If the symmetry of the crystal is also lowered to the monoclinic one, half of the JT centers in the common position is not connected by any symmetry elements with the other part. The distortions of the surrounding of the JT ions, which belong to different groups, provide different energies of non-equivalent rare-earth ions. 

So, when the temperature is decreased the electronic mode shifts towards higher energies and splits into two modes, $\omega_{e1}^{IR}$ and $\omega_{e2}^{IR}$, with energies of 22 and 27~cm$^{-1}$ of different intensities, respectively. The temperature independent sharp peak at 17.5~cm$^{-1}$ marked with $\omega_{ph}^{IR}$ is the phonon mode which corresponds to the shear lattice vibrations \cite{Pop}. At high temperatures the electron excitation (as a broad line) basically superimposes on the phonon excitation. We think one can suppose that the infrared active phonons of that nature are not ``triggers'' of the phase transition. More preferable is the assumption about the participation in the mechanism of the JT transition of the $A_u$ low-frequency phonons connected with the weak turns of the tetrahedral anions of (MoO$_4$)$^{2-}$,  which deform the local environment of the Dy$^{3+}$ ions. Those excitations are non-active either in the FIR or in the Raman spectra, and their low-frequency nature can be manifested only with the help of the modelling. 

\begin{figure}
\begin{center}
\includegraphics[width=0.34\textwidth]{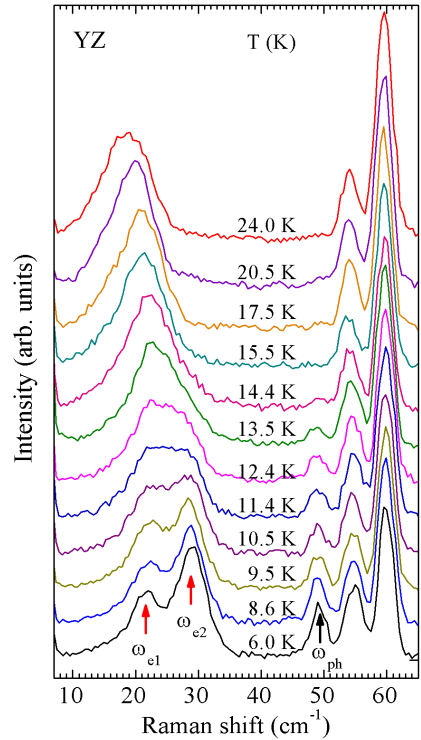}
\end{center}
\caption{(Color online) The temperature behavior of the low-frequency part of the Raman spectrum of KDy(MoO$_4$)$_2$. The geometry of the experiment is $Z(YZ)Y$ ($B_{3g}$ symmetry). Black arrow shows the additional phonon line, and red arrows show the low-energy electron transitions (related to the ones between the levels of the main multiplet  $^6H_{15/2}$ in the high-temperature phase) of the Dy$^{3+}$ ion. The spectral resolution is 3~cm$^{-1}$.}
\label{fig2}
\end{figure}
 
Fig.~2 shows the temperature evolution of the Raman spectra between 6 and 24~K for the $Z(YZ)Y$ geometry (experiments have been performed for other geometries too), which probes $B_{3g}$ modes. There are three peaks with the energies below 70~cm$^{-1}$ at 24~K. The broad peak at 18~cm$^{-1}$ corresponds to the electronic excitation within $^6H_{15/2}$ multiplet and the peaks at 55 and 60~cm$^{-1}$ are related to phonons \cite{Pes}. Upon cooling below 14.5~K the new phonon mode appears at 49.2~cm$^{-1}$ (shown by the black arrow). The onset of the phonon modes below the critical temperature points towards the multiplication of the unit cell because in the rhombic phase (above 14.5~K) the degeneracy of the phonons is already removed. A few more new phonons were observed in the low-temperature phase \cite{Pes}. The electronic mode with the energy of 18~cm$^{-1}$ at 24~K shifts upon cooling to higher energies and splits to two peaks at 21.1 and 28.8~cm$^{-1}$ (shown by red arrows). Two other polarizations show the analogous behavior \cite{Pes}.  

\begin{figure}
\begin{center}
\includegraphics[width=0.48\textwidth]{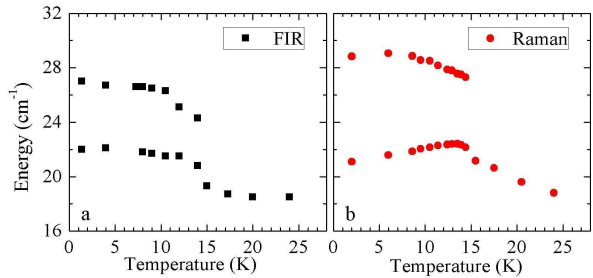}
\end{center}
\caption{(Color online) The temperature dependence of the low-energy electronic excitations of Dy$^{3+}$ ions in KDy(MoO$_4$)$_2$ in the vicinity of phase transition(s). The left panel: the FIR measurements; the right panel: the Raman measurements. Raman data for 2~K are used from Ref.~\cite{Pes}.}
\label{fig3}
\end{figure}

Note, that in \cite{Mih} the observed excitation of the electronic nature about 21$\div$22~cm$^{-1}$ was interpreted as a soft mode, however without supporting  experimental data. That softening of the mode is directly connected to the equivalence of two ordered quadrupolar sublattices \cite{Mih}. Instead, in our experiment it is shown that both low-temperature lines do {\em not} shift to zero. Moreover, Raman experiments \cite{Pes} of various polarizations and at various frequency ranges do not reveal any sorts of “soft” modes. 
 
Fig.~3 summarizes the results of our Raman and FIR studies. It shows the temperature dependence of the energies of electronic excitations. Below the phase transition the gap between two lowest electron energies increases. The mode splitting is clearly visible in both experiments. We assign the splitting of the electronic mode to the increase of the non-equivalent positions of the Dy$^{3+}$ ions which means that the symmetry in the low temperature phase is not higher than monoclinic in the volume-doubled elementary cell of the rhombic phase \cite{suppl,polar}. We address the difference in energies of the electronic excitations in the Raman and FIR measurements to eight non-equivalent positions of Dy$^{3+}$ ions. The interaction between those ions additionally splits the electronic levels. The schematic picture of the low-temperature phase unit cell is shown in Fig.~4. There are eight rare-earth ions which occupy the common crystallographic positions. They consist of two groups, each of which includes four ions. Those four ions are connected with each other by the inversion operation, however no symmetry operation exists which transfer one group to the other one. 

\begin{figure}
\begin{center}
\includegraphics[width=0.22\textwidth]{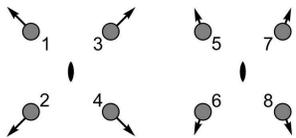}
\end{center}
\caption{The schematic picture of the distorted Dy$^{3+}$ ions in KDy(MoO$_4$)$_2$ below the phase transition to the ferriquadrupolar phase.}
\label{fig4}
\end{figure}

Obviously, such a behavior evidences \cite{magn} the onset of the ferriquadrupolar electron ordering, in which two ordered sublattices of orbital quadrupoles are non-equivalent \cite{af}. It is accompanied with non-equivalent distortions of two groups of of the oxygen polyhedrons surrounding two groups   of Dy$^{3+}$ ions, see Fig.~4. To explain the observed behavior of the interacting via the JT effect electron and elastic subsystems of KDy(MoO$_4$)$_2$ we develop the approach proposed in \cite{JT,Ell,Leask}, modifying it according to the observed in our experiment non-equivalence of two quadrupolar sublattices. Such an approach has been successfully applied for the description of the features of the infrared and optical spectroscopy in rare-earth based JT systems \cite{JT,Ell,Leask,Mih,Kolod}. 

At high temperatures the first excited doublet of Dy$^{3+}$ is situated 18~cm$^{-1}$ above the ground state doublet due to the crystalline electric field, hence for the description of low-energy properties of the system we limit ourselves with those two doublets \cite{JT,Ell} (often called Ising pseudo-spins). Quadrupolar moments of Dy$^{3+}$, related to those doublets \cite{Ell}, interact with each other via phonons with the long-range Ising type of the effective coupling \cite{JT}, yielding quadrupolar ordering at low temperatures. That interaction is taken into account within the mean-field theory for the quadrupolar ordering \cite{Leask}. In our calculations the value of the interaction between quadrupolar moments, belonging to the same sublattice, is taken to be equal to the value of the interaction between quadrupoles, belonging to different sublattices (however with the different sign). It has been used to avoid the additional magnetic field-induced ordered phase, analogous to the spin-flop phase of antiferromagnets, not observed in the experiment \cite{Kolod} (that phase is necessarily predicted by the mean-field theory as in \cite{Leask}, if those interactions have different values). 

Within such an approach the low-energy part of the free energy per site of quadrupolar moments (Ising pseudo-spins) coupled to the elastic subsystem in the simplest approximation can be written as $F=- (k_BT/2) (\ln [2\cosh (A_+/k_BT)] + \ln [2\cosh (A_-/k_BT)] ) +Cx^2/2$, where $T$ is the temperature, $k_B$ is the Boltzmann constant, $A_{\pm}= [(\lambda L)^2(1\pm x)^2+\varepsilon^2]^{1/2}$ are the renormalized (half)splittings of energies of quadrupolar moments (namely transitions between those low-energy electron states were observed in our spectroscopic experiments), $\lambda$ is the parameter of the effective inter-quadrupole Ising-like interaction, $\varepsilon$ is the crystalline electric field, which determines the high-temperature (half)splitting of 18~cm$^{-1}$, and $L$ is the order parameter \cite{Leask,Kolod}. Two first terms are related to the energy gain of the quadrupolar electron subsystem due to the non-equivalence of quadrupoles due to the nonzero parameter of the non-equivalence $x$. On the other hand, the last term describes the energy loss of the elastic subsystem due to the non-equivalence of strains of two electron sublattices (cf. Fig.~4), with $C$ being related to the elastic modulus. For $x=0$, i.e., for equivalent quadrupolar sublattices (the antiferroquadrupolar ordering), the order parameter $L$ is the solution of the self-consistency equation $A=\lambda \tanh (A/k_BT)$, where $A= [(\lambda L)^2 + \varepsilon^2]^{1/2}$, see \cite{suppl}. 

In the ground state the order parameter $L$ is equal to $[1-(\varepsilon/\lambda)^2]^{1/2}$, and it monotonically decreases to zero at $T = T_D = \lambda/k_B$. That type of ordering exists only for $\lambda \ge  \varepsilon$. Then, to describe the non-equivalence of the quadrupolar sublattices, i.e., the ferriquadrupolar ordering, we minimize the free energy $F$ with respect to $x$. We obtain the equation for the determination of $x$:
\begin{equation}
\frac {2Cx}{(\lambda L)^2}=\frac{(1+x)}{A_+}\tanh \left(\frac {A_+}{k_BT}\right) -\frac{(1-x)}{A_-}\tanh \left(\frac{A_-}{k_BT}\right). 
\end{equation}
Obviously for $T \gg A_{\pm}/k_B$ only trivial solution $x=0$ exists. We can see that a nontrivial solution $x\ne 0$, which describes the ferriqudruplolar ordering with non-equivalent quadrupole sublattices $L(1\pm x)$ exists only for $T <T_D$, where $L\ne 0$. Unfortunately, it is impossible to find the analytic solution to Eq.~(1). Our analysis of the numerical solution shows that for given values of $\lambda$ and $\varepsilon$ the nontrivial solution for $x \ne 0$ exists, however only for some range of values of $C$, which is reasonable for KDy(MoO$_4$)$_2$. Depending on the values of the used parameters, the model can describe the single transition from the disordered high temperature phase to the ferriquadrupolar ordered phase directly. The other possibility is via two phase transitions, first from the disordered phase with $L=0$ to the antiferroquadrupolar one (with equivalent sublattices, $L \ne 0$, $x=0$), and then to the ferriquadrupolar phase with non-equivalent sublattices  $L \ne 0$, $x \ne 0$. 

In summary, using the FIR and Raman spectroscopies we have directly observed the spontaneous low-temperature ferriquadrupolar ordering of quadrupolar moments of the rare-earth ions in the rare-earth-based compound KDy(MoO$_4$)$_2$. Our FIR and Raman studies also manifest that such ordering is accompanied with the new optical phonon mode, related to the non-equivalent distortion of rare-earth ions, belonging to observed non-equivalent quadrupolar sublattices. The developed mean-field theory explains the main mechanism of the transition(s) to the low-temperature ferriquadrupolar phase, observed in our experiments. Four non-equivalent Ising-like magnetic centers observed in the external magnetic field \cite{Bag} in the low-temperature phase of KDy(MoO$_4$)$_2$ also support our conclusion about the ferriquadrupolar character of the low-temperature phase of the electron subsystem.

We acknowledge the support of the HFML, member of the European Magnetic Field Laboratory (EMFL). We thank V.~Kurnosov for useful discussions.


\begin{thebibliography}{99}
\bibitem{not} Notice that for rear-earth systems due to the strong spin-orbit coupling spin and orbital degrees of freedom are strongly coupled.
\bibitem{polar} The results of Raman studies for different polarizations of light \cite{Pes} also agree with our conclusions.
\bibitem{magn} There is no magnetic ordering in KDy(MoO$_4$)$_2$ in the considered range of the temperature. The magnetic ordering takes place below 
approx. 1.1~K (see \cite{Cooke,Vit}), and has nothing in common with the observed below appr. 15~K anomalies, related to the orbital quadrupolar ordering (and simultaneous distorsive ordering, characteristic to the co-operative Jahn-Teller effect) in KDy(MoO$_4$)$_2$, see \cite{Leask,Nest, Kharch,Cooke,Mih,Vit,Kut,Kolod}. The nature of the low-temperature magnetic ordering is related to magnetic dipole-dipole interactions, which are too weak to explain the observed features in the considered in our study temperature range.
\bibitem{af} The ferroquadrupole ordering is known to be excluded for KDy(MoO$_4$)$_2$ \cite{Leask,Nest, Kharch,Cooke,Mih,Vit,Kut,Kolod}. As for the antiferroquadrupolar ordering: Equivalent quadrupole sublattices would yield softening of the lower electron mode at the critical temperature \cite{Mih}. Such a softening is obviously absent (see Fig.~3) in both observed FIR and Raman data (also for other polarizations of light). The model with non-equivalent sublattices, i.e., the model of the ferriquadrupolar ordering, see below, does not reveal such a softening.
\end{thebibliography}
\end{document}